\documentclass[preprint2,numberedappendix]{emulateapj-rtx4}
\usepackage{bm,color}
\newcommand{\Fig}[1]{Figure~\ref{#1}}
\newcommand{\Figp}[2]{Figure~\ref{#1}({#2})}
\newcommand{\bra}[1]{\langle #1\rangle}
\newcommand{\m}{\,{\rm m}}
\newcommand{\s}{\,{\rm s}}
\newcommand{\Mm}{\,{\rm Mm}}
\newcommand{\Eq}[1]{Equation~(\ref{#1})}
\newcommand{\Eqs}[2]{Equations~(\ref{#1}) and~(\ref{#2})}
\newcommand{\EEq}[1]{Equation~(\ref{#1})}
\def\half{{\textstyle{1\over2}}}
\newcommand{\EK}{E_{\rm K}}
\newcommand{\EM}{E_{\rm M}}
\newcommand{\HM}{H_{\rm M}}
\newcommand{\HC}{H_{\rm C}}
\newcommand{\G}{\,{\rm G}}
\newcommand{\g}{\,{\rm g}}
\newcommand{\cm}{\,{\rm cm}}

\def\Brms{B_{\rm rms}}
\newcommand{\xx}{\bm{x}}
\newcommand{\kk}{\bm{k}}
\newcommand{\KK}{\bm{K}}
\newcommand{\BB}{\bm{B}}
\newcommand{\AAA}{\bm{A}}
\newcommand{\nullvector}{\bm{0}}

\newcommand{\xxi}{\bm{\xi}}
\newcommand{\yjour}[4]{ #1, {#2}, {#3}, #4}
\newcommand{\yasr}[3]{ #1, {Adv. Spa. Res.,} {#2}, #3}
\newcommand{\yapj}[3]{ #1, {ApJ,} {#2}, #3}
\newcommand{\yapjl}[3]{ #1, {ApJL,} {#2}, #3}
\newcommand{\yan}[3]{ #1, {Astron. Nachr.,} {#2}, #3}
\newcommand{\yana}[3]{ #1, {A\&A,} {#2}, #3}
\newcommand{\ysov}[3]{ #1, {Sov. Astron.,} {#2}, #3}
\newcommand{\ygafd}[3]{ #1, {Geophys. Astrophys. Fluid Dyn.,} {#2}, #3}
\newcommand{\ysph}[3]{ #1, {Solar Phys.,} {#2}, #3}
\newcommand{\yprl}[3]{ #1, {PRL,} {#2}, #3}
\newcommand{\ypre}[3]{ #1, {PRE,} {#2}, #3}
\newcommand{\ypf}[3]{ #1, {Phys. Fluids,} {#2}, #3}

\newcommand{\papj}[2]{ #1, {ApJ}, in press, arXiv:#2}

\begin{document}
\preprint{NORDITA-2018-35}  

\title{Solar Kinetic Energy and Cross Helicity Spectra}

\author{Hongqi Zhang$^1$ and Axel Brandenburg$^{2,3,4,5}$}
\affil{
$^1$Key Laboratory of Solar Activity, National Astronomical Observatories, Chinese
Academy of Sciences, Beijing 100012, China, \\
$^2$Nordita, KTH Royal Institute of Technology and Stockholm University,
Roslagstullsbacken 23, SE-10691 Stockholm, Sweden,\\
$^3$JILA and Department of Astrophysical and Planetary Sciences,
University of Colorado, Boulder, CO 80303, USA\\
$^4$Department of Astronomy, AlbaNova University Center,
Stockholm University, SE-10691 Stockholm, Sweden\\
$^5$Laboratory for Atmospheric and Space Physics,
University of Colorado, Boulder, CO 80303, USA
}

\submitted{\today,~ $ $Revision: 1.55 $ $}

\begin{abstract}
{We develop a formalism that treats the calculation of solar kinetic
energy and cross helicity spectra in an equal manner to that of magnetic
energy and helicity spectra.
The magnetic helicity spectrum is shown to be equal to the vertical
part of the current helicity spectrum divided by the square of the wavenumber.
For the cross helicity, we apply the recently developed two-scale
approach globally over an entire active region to account for the sign
change between the two polarities.
Using vector magnetograms and Dopplergrams of NOAA~11158 and 12266,
we show that kinetic and magnetic energy spectra have similar slopes at
intermediate wavenumbers, where the contribution from the granulation
velocity has been removed.}
At wavenumbers around $0.3\Mm^{-1}$, the magnetic helicity is found to
be close to its maximal value.
The cross helicity spectra are found to be within about 10\% 
of the maximum possible value.
{Using the two-scale method for NOAA~12266, the global cross helicity spectrum is found to
be particularly steep, similarly to what has previously been found in
theoretical models of spot generation.}
In the quiet Sun, by comparison, the cross helicity spectrum
is found to be small.
\end{abstract}

\email{hzhang@bao.ac.cn}

\section{Introduction}

The photospheric plasma is thought to be in a state of fully developed
magnetohydrodynamic turbulence at high magnetic Reynolds numbers.
The spectral slope of the turbulent velocity field is believed to be of the order of
$-5/3$, as theoretically proposed by \cite{Ko41} and \cite{Obukhov41},
and as is also expected from the theory of nonhelical hydromagnetic
turbulence when the magnetic field is moderately strong and therefore
noticeably anisotropic \citep{GS95}.
For decaying turbulence, on the other hand, \cite{Lee2010} found that the scaling depends on the
field strength, and a shallower Iroshnikov--Kraichnan $k^{-3/2}$ spectrum
\citep{Iro63,Kraichnan65} occurs for weaker fields and a steeper $k^{-2}$
weak-turbulence spectrum for stronger fields \citep{Gal00,BKT15}; see the
reviews by \cite{BS05} and \cite{BN11} for a discussion of the respective
phenomenologies in the three cases.
The observations of magnetic and velocity fields in the solar atmosphere
provide a window to analyze solar hydromagnetic turbulence through
their power spectra and compare with earlier work \citep{Abr05, AY10,
Sten12, ZhaoChou13}.

The technique used to obtain the scale dependence of magnetic helicity
through observations is reminiscent of that of \cite{MGS82}, who made
the assumption of isotropy to express the Fourier transform of the
two-point correlation tensor of the magnetic field in terms of the
Fourier transforms of the magnetic field.
Their approach made use of one-dimensional spectra obtained from
time series of all three magnetic field components
and was applied to in situ measurements in the solar wind.
The Taylor hypothesis \citep{Taylor38} was used to relate
the two-point correlation function in time to one in space.
In the work of \cite{Zhang2014, Zhang2016}, again the assumption of
isotropy was made, but a full two-dimensional array of magnetic field
vectors was used, so the Taylor hypothesis was not invoked. They applied
this technique to a number of active regions to determine magnetic energy
and helicity spectra and their change with time.
The current helicity spectrum was estimated from the magnetic
helicity spectrum under the assumption of isotropy, and its modulus
showed a $k^{-5/3}$ spectrum at intermediate wavenumbers.
A similar power law is also obtained for the magnetic energy spectrum.
These are largely consistent with expectations from the turbulence
simulations discussed by \cite{BS05}.

In this letter we compare power spectra of the magnetic field
with those of the velocity field inferred from photospheric
vector magnetograms and Dopplergrams in solar active regions NOAA~11158,
12266, and the quiet Sun.
In addition to the kinetic energy spectrum, we also compute cross
helicity spectra.
Cross helicity has been determined previously using both theory
\citep{Pip11,RKB11,BR13,Yok13} and observations \citep{KPZ07,RKS12,Zha14},
and it may play a direct role in the production of active regions
\citep{BGJKR14}.
However, cross helicity spectra have previously only been obtained from theory.

\section{Basic formalism}

We begin by reviewing briefly the method of \cite{Zhang2014}.
They introduced the two-point correlation tensor of the magnetic field,
$\langle B_i({\xx},t) B_j({\xx}+\xxi,t)\rangle$, and write
its Fourier transform with respect to $\xxi$ as
\begin{equation}
\left\langle\tilde{B}_i({\kk},t)\tilde{B}_j^*\!({\kk}',t)\right\rangle
=\Gamma_{ij}({\kk},t)\delta^2({\kk}-{\kk}'),
\end{equation}
where the tildes indicate Fourier transformation, i.e.,
$\tilde{B}_i({\kk},t)=\int B_i({\xx},t)\,e^{i{\kk}\cdot{\xxi}}d^2\xi$,
and the asterisk denotes complex conjugation.
Under the assumption of isotropy, the spectral correlation tensor
$\Gamma_{ij}({\kk},t)$ can be written as
\begin{equation}
\Gamma_{ij}({\kk},t)=\frac{2E_{\rm M}(k,t)}{4\pi k}(\delta_{ij}-\hat{k}_i\hat{k}_j)
+\frac{{\rm i}H_{\rm M}(k,t)}{4\pi k}\varepsilon_{ijk}k_k,\label{eq:helispec5}
\end{equation}
where $E_{\rm M}(k,t)$ and $H_{\rm M}(k,t)$ are the shell-integrated
magnetic energy and helicity spectra, respectively, $\hat{{\kk}}=\kk/k$
is the unit vector of $\kk$, and $k=(k_x^2+k_y^2)^{1/2}$ is the wavenumber.
The spectra are normalized such that $\int E_{\rm M}\,{\rm d}k=\langle{\BB}^2\rangle/2$
and $\int H_{\rm M}\,{\rm d}k=\langle{\AAA}\cdot{\BB}\rangle$, where ${\AAA}$
is the magnetic vector potential with ${\BB}=\bm\nabla\times{\AAA}$.
The two spectra can also be computed as \citep[cf.][]{BN11}
\begin{eqnarray}
E_{\rm M}(k)&=&\;\half\!\!\!\!\!\!\!\!\sum_{k_- < |{\kk}|\leq k_+} \!\!\!\!\!\! |\tilde{\BB}({\kk})|^2,\label{Evec}\\
H_{\rm M}(k)&=&\;\half\!\!\!\!\!\!\!\!\sum_{k_- < |{\kk}|\leq k_+} \!\!\!\!\!\!
(\tilde{\AAA}\cdot\tilde{\BB}^\ast+\tilde{\AAA}^\ast\cdot\tilde{\BB}),
\label{EHvec}
\end{eqnarray}
where $k_\pm=k\pm\delta k/2$ and $\delta k=2\pi/L$ is the wavenumber
increment and also the smallest wavenumber in the plane $L^2$ with $L$
being the size of the magnetograms.
Following common convention, the magnetic energy density is measured
in $\G^2$, so the units of the spectrum $E_{\rm M}(k)$ are $\G^2\cm$
\citep{AY10a}.
To compute the magnetic helicity spectrum, \cite{Zhang2014} used the expression
\begin{eqnarray}
kH_M(k,t)&=&4\pi k\,\mbox{Im}\left\langle\cos\phi_k\Gamma_{yz}
-\sin\phi_k\Gamma_{xz}\right\rangle_{\phi_k},
\label{kH_M}
\end{eqnarray}
where we have defined the polar angle in wavenumber space
so that $k_x=k\cos\phi_k$ and $k_y=k\sin\phi_k$.
The angle brackets with subscript $\phi_k$
denote averaging over annuli in wavenumber space.
{Note that only the $xz$ and $yz$ components enter, so
\Eq{kH_M} becomes
\begin{eqnarray}
kH_M(k,t)&=&4\pi k\,\mbox{Im}\left\langle
\left(k_x\tilde{B}_y-k_y\tilde{B}_x\right)
\tilde{B}_\|^\ast\right\rangle_{\phi_k}.
\end{eqnarray}
Thus, by introducing
\begin{eqnarray}
\tilde{A}_\|=(-ik_x\tilde{B}_y+ik_y\tilde{B}_x)/k^2
\equiv\tilde{J}_\|/k^2,
\end{eqnarray}
with $\tilde{J}_\|$ being the Fourier transform of
$J_\|\equiv\partial_x B_y-\partial_y B_x$, we can relate
$\HM(k,t)$ to the vertical part (indicated by $\|$) of the current helicity spectrum,
$\HC(k,t)=k^2\HM(k,t)$, which was already used in \cite{Zhang2014}.
Therefore, instead of working with} Equation~(\ref{EHvec}),
we compute from now on
\begin{eqnarray}
\label{}
E_{\rm M}(k)&=&\;\half\!\!\!\!\!\!\!\!\sum_{k_- < |{\kk}|\leq k_+} \!\!\!\!\!\! |\tilde{B}_\|({\kk})|^2,\\
H_{\rm M}(k)&=&\;\half\!\!\!\!\!\!\!\!\sum_{k_- < |{\kk}|\leq k_+} \!\!\!\!\!\!
(\tilde{A}_\|\tilde{B}_\|^\ast+\tilde{A}_\|^\ast\tilde{B}_\|),
\end{eqnarray}
which is equivalent to \Eqs{Evec}{EHvec}.
Note that $E_{\rm M}(k)$ and $H_{\rm M}(k)$ satisfy the realizability condition
$2E_{\rm M}\ge k|H_{\rm M}|$, which is also the reason why we always plot the
scaled combinations $2E_{\rm M}(k)$ and $kH_{\rm M}(k)$. This allows us to
judge how close to fully helical the magnetic field is at each wavenumber.

An analysis similar to that of the magnetic field can also be done for
the velocity.
Only the Doppler velocity field can readily be observed.
Thus, we compute vertical kinetic energy and cross helicity spectra as
\begin{eqnarray}
E_{\rm K}(k)&=&\;\half\!\!\!\!\!\!\!\!\sum_{k_- < |{\kk}|\leq k_+} \!\!\!\!\!\! |\tilde{v}_\|({\kk})|^2,\\
H_{\rm X}(k)&=&\;\half\!\!\!\!\!\!\!\!\sum_{k_- < |{\kk}|\leq k_+} \!\!\!\!\!\!
(\tilde{v}_\|\tilde{B}_\|^\ast+\tilde{v}_\|^\ast\tilde{B}_\|).
\label{EKEMHX}
\end{eqnarray}
Defining $q\equiv\sqrt{4\pi\rho_0}$, the realizability condition reads
\begin{equation}
qE_{\rm K}(k)+E_{\rm M}(k)/q \ge \, |H_{\rm X}(k)|,
\label{Elimit}
\end{equation}
where we assume $\rho_0=10^{-6}\g\cm^{-3}$ for the density and
ignore density fluctuations.
As $v_\|$ is measured in ms$^{-1}$ and $B_\|$ in G, we have
$q=100\cm\m^{-1}\sqrt{4\pi\rho_0}$.
We determine $E_{\rm K}(k)$ and $H_{\rm X}(k)$ to study the spectral
distribution of the line-of-sight velocity, and its relationship with
that of the magnetic field.

\begin{figure}
\begin{center}
\includegraphics[width=.23\textwidth]{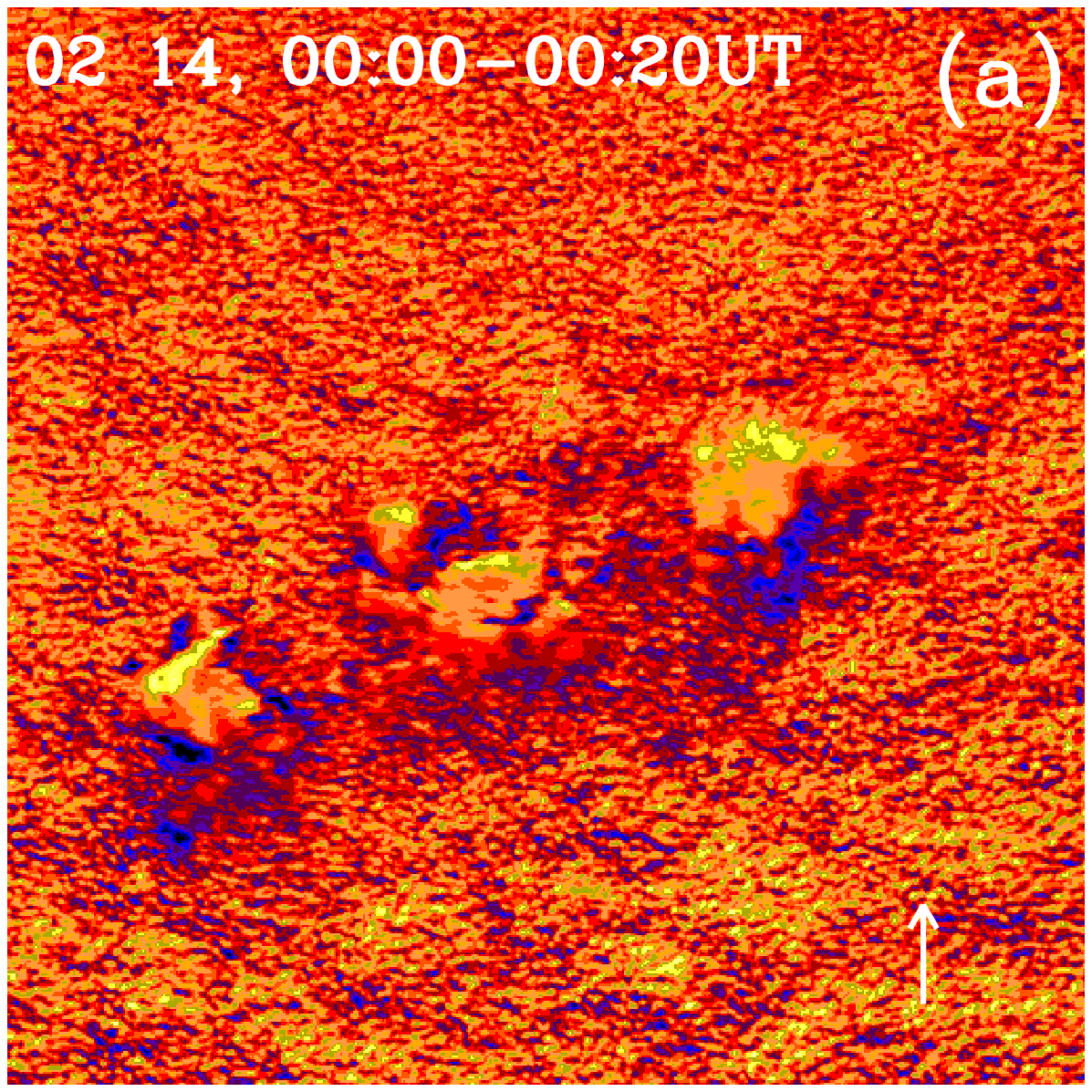}
\includegraphics[width=.23\textwidth]{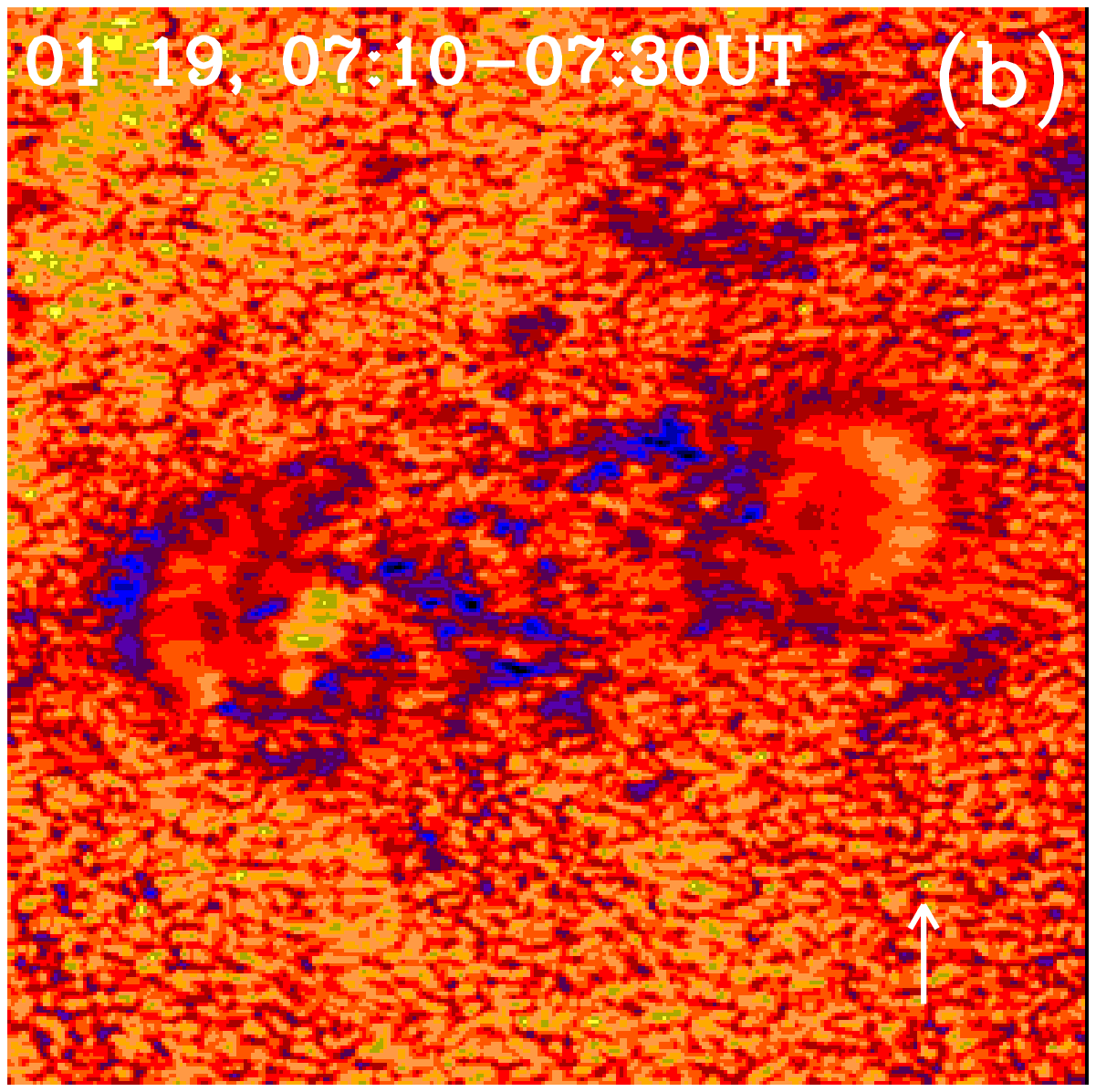}
\vspace*{0.1mm}
\includegraphics[width=.23\textwidth]{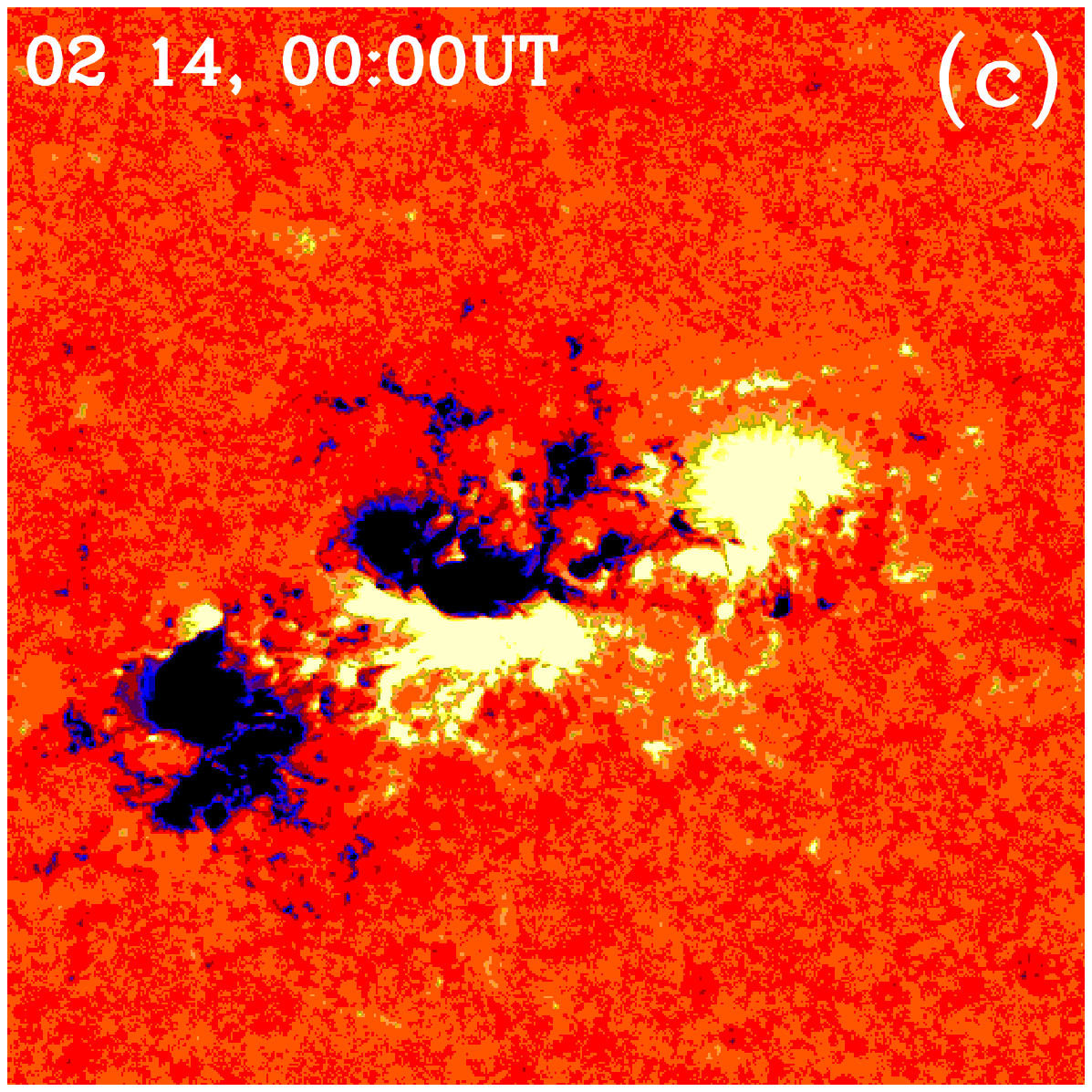}
\includegraphics[width=.23\textwidth]{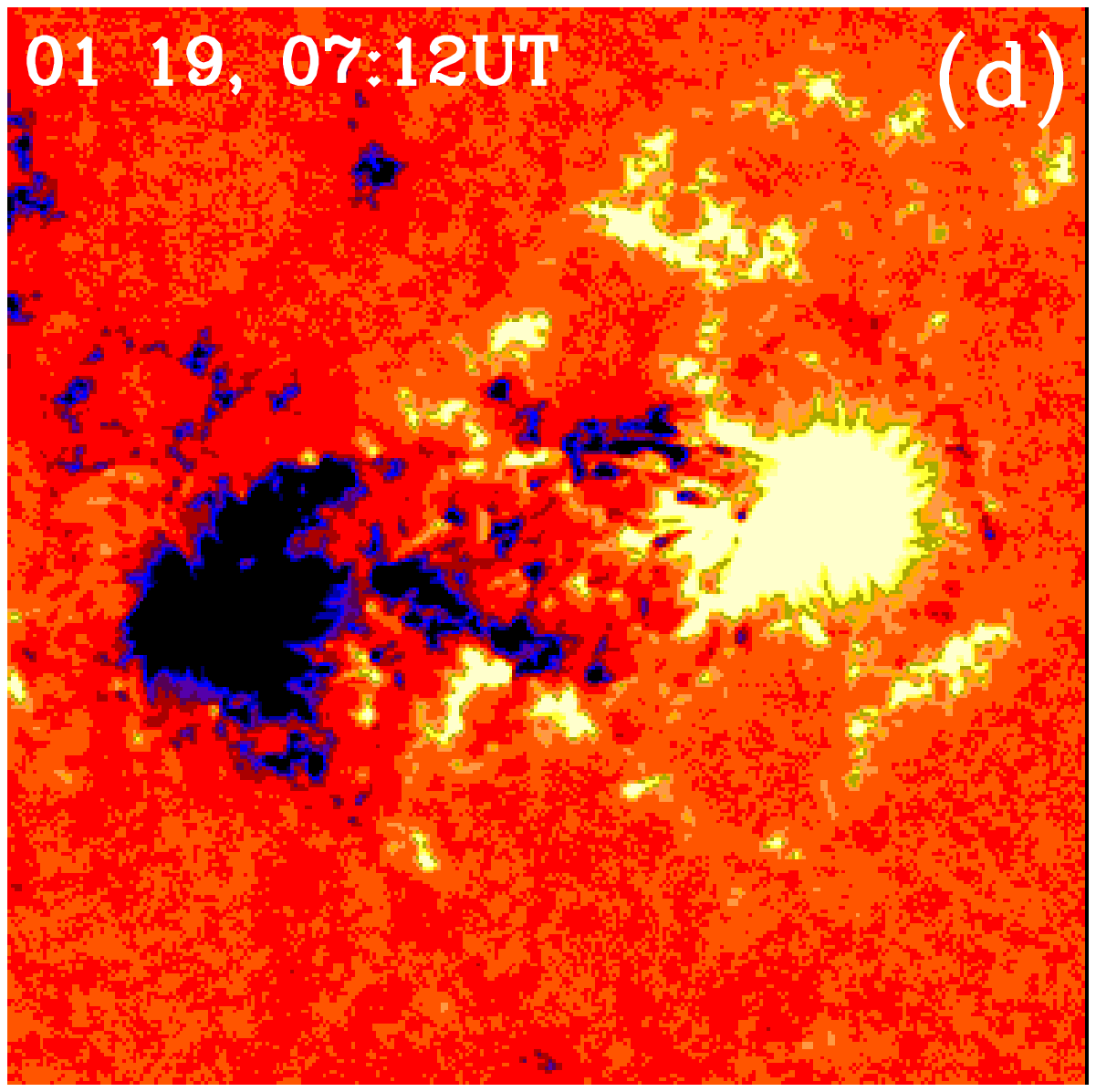}
\vspace*{-1mm}
\includegraphics[width=.23\textwidth]{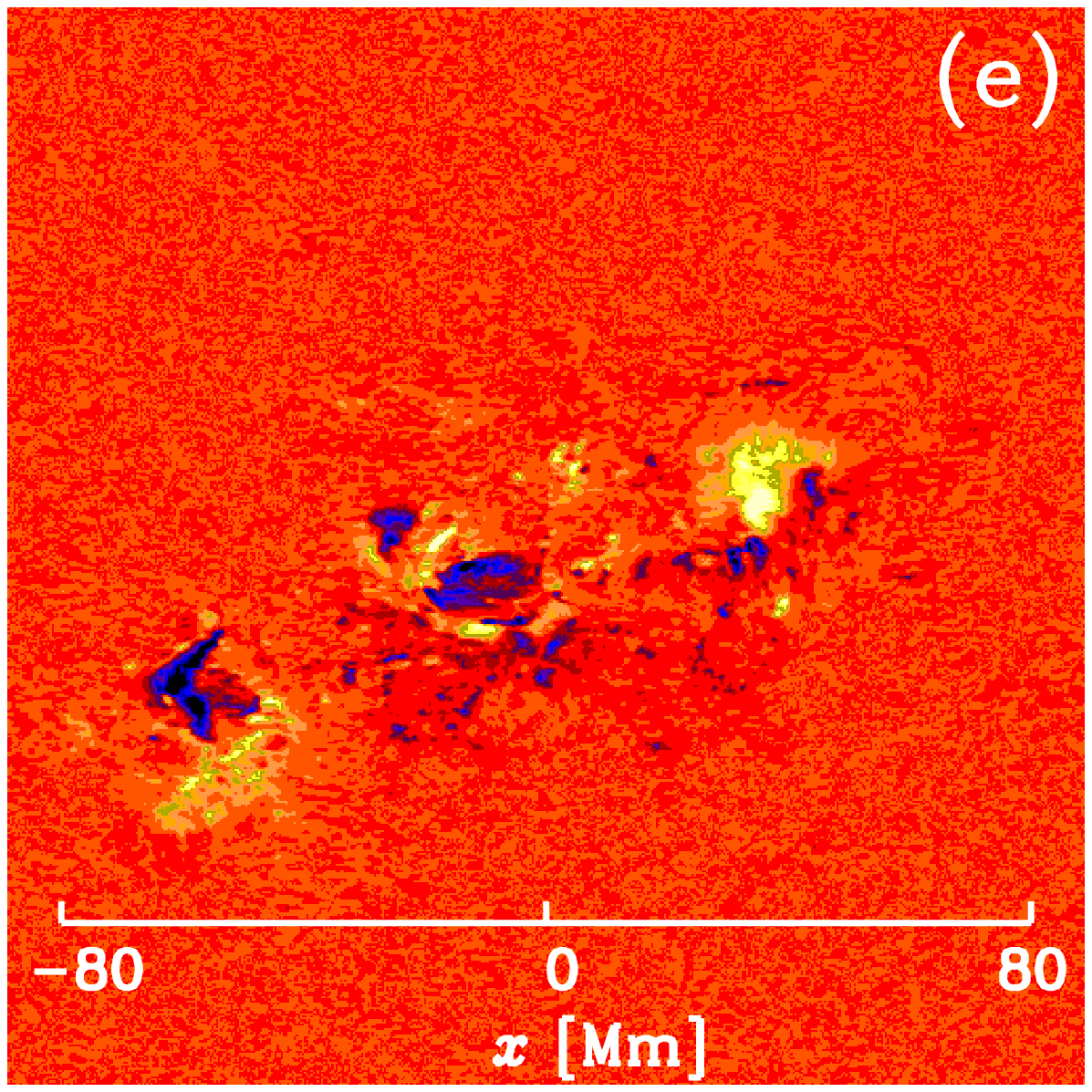}
\includegraphics[width=.23\textwidth]{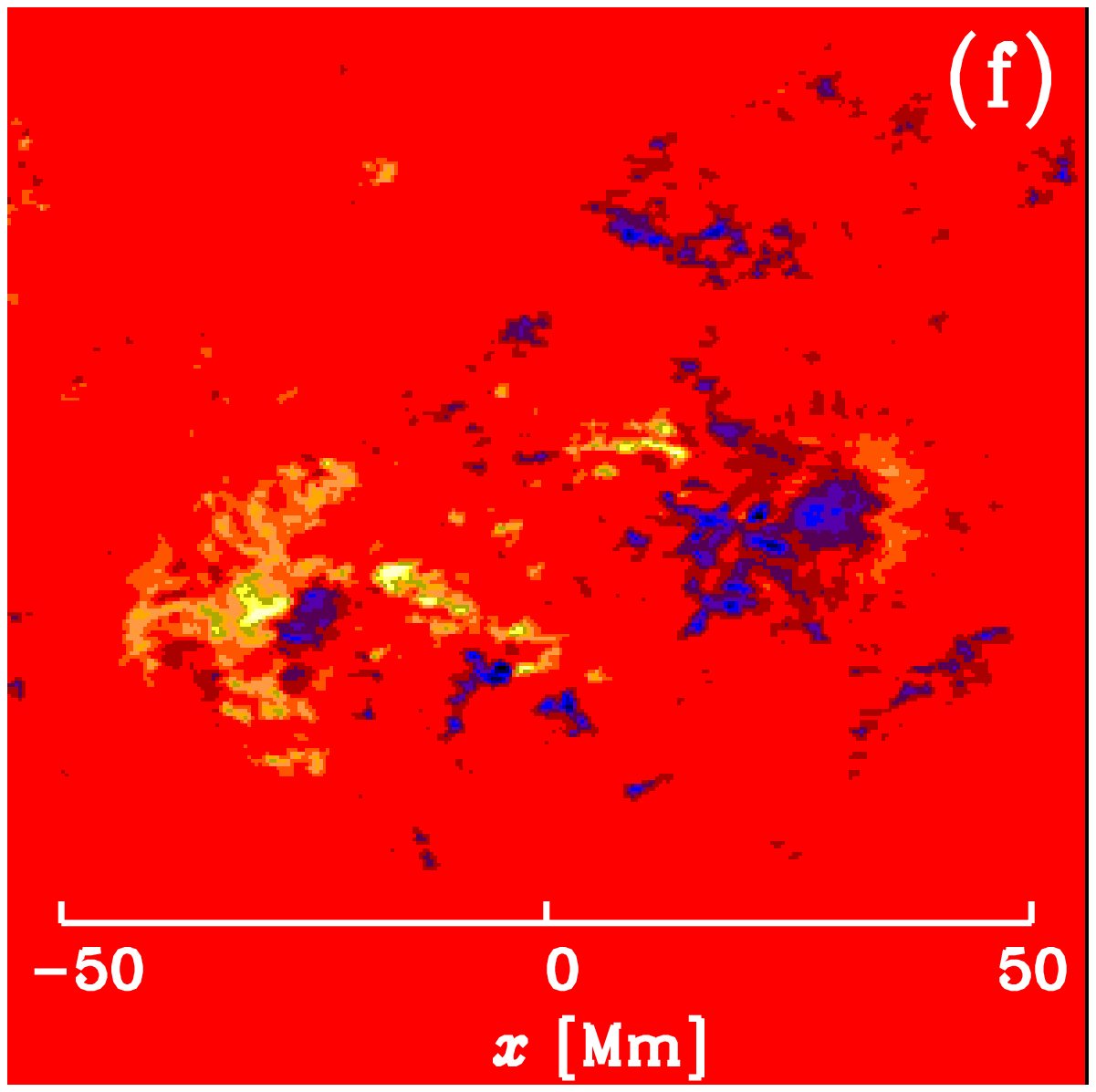}
\end{center}
\caption{Doppler velocity field $v_\|$ (top),
longitudinal magnetic field $B_\|$ (middle),
and their product $v_\|B_\|$ (bottom), for 
active regions NOAA~11158 (left, field of view $256''\times256''$)
on 2011 February 14 and NOAA~12266 (right, field of view $150''\times150''$)
on 2015 January 19.
Yellow (blue) shades show positive (negative), corresponding to
the upward (downward) directions.
\label{fig:corrmagheliveloa}
}\end{figure}

\begin{figure*}
\begin{center}
\hspace*{-10mm}
\includegraphics[width=80mm]{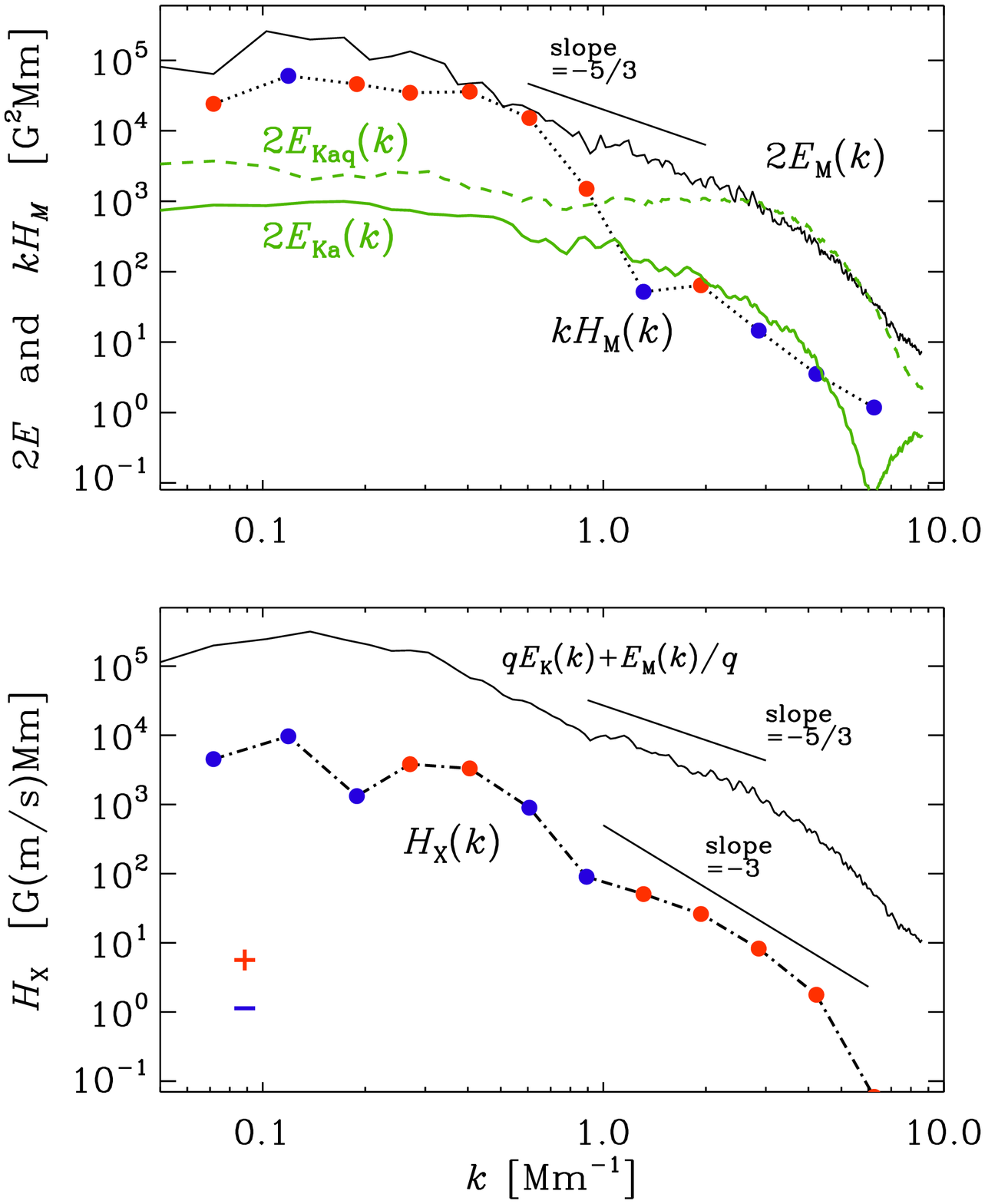}
\hspace*{2mm}
\includegraphics[width=80mm]{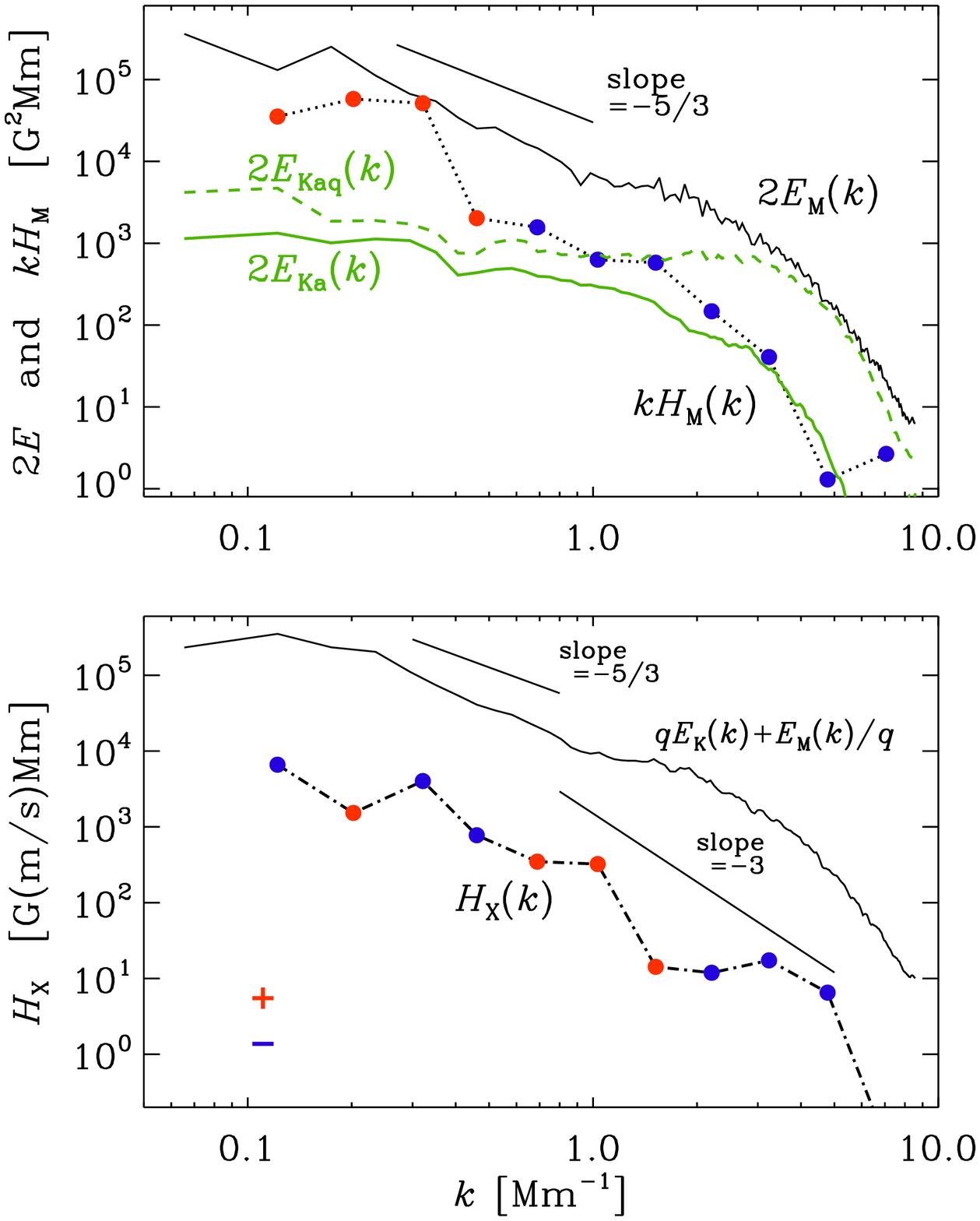}
\end{center}
\caption{{The upper panels show spectra}
of magnetic energy $E_{\rm M}(k)$ (black solid
lines), normalized magnetic helicity $kH_{\rm M}(k)$ (black dotted lines;
{red and blue symbols denote positive and negative values, respectively})
and kinetic energy $E_{\rm Kaq}(k)$ (green dashed lines for
kinetic energy of the quiet Sun) and $E_{\rm Ka}(k)$ (green {solid}
lines for kinetic energy related to magnetic features only) in active
region NOAA~11158 (left) and NOAA~12266 (right).
{The lower panels show $qE_{\rm K}(k)+E_{\rm M}(k)/q$ (solid lines)
and $H_{\rm X}(k)$ (dashed-dotted lines; red and blue symbols denote positive
and negative values, respectively).}
\label{fig:corrmaghelivelob}
}\end{figure*}

For the cross helicity, there is a particular problem when considering
bipolar active regions.
We expect the cross helicity to be proportional to the mean ambient
magnetic field \citep{RKB11}.
It will therefore have contributions of opposite signs from the two
poles of a bipolar
region, leading to cancelation; see Figure~\ref{fig:corrmagheliveloa}
for visualizations of $v_\|$, $B_\|$, and $v_\|B_\|$ for
the active regions NOAA~11158 and 12266.

NOAA~12266 has two clearly separated poles.
A suitable technique to obtain a spectrum encompassing the entire
bipolar region is the two-scale approach of \cite{BPS17},
who applied it to measuring magnetic helicity for the entire solar disk,
taking the systematic sign change across the equator into account;
see also \cite{Singh18}.
This technique allows us to incorporate the sign change as a sinusoidal
modulation proportional to $\sin\KK\cdot\xx$ with wavevector $\KK$ and,
in principle, arbitrary phase shifts, which are not considered here.
\EEq{EKEMHX} then becomes 
\begin{equation}
H_{\rm X}(\KK,k)=\!\!\!\!\!\!\!\!
\sum_{k_- < |{\kk}|\leq k_+}\!\!\!\!\!\!\!\!
\tilde{v}_\|(\kk+\KK/2)\tilde{B}_\|^\ast(\kk-\KK/2),\;
\label{EKEMHXdx}
\end{equation}
which is complex, and its real part agrees with \Eq{EKEMHX} for
$\KK=\nullvector$.
For a bipolar region aligned in the $x$ direction with an approximate
separation $d$, we have $\KK=(\pi/d,0)$.
Analogous to \cite{BPS17}, the relevant spectrum is then
$-{\rm Im}\,H_{\rm X}(\KK,k)$.
We return to this below when discussing concrete examples.

\section{Comparison of the Spectra}

Figure~\ref{fig:corrmagheliveloa} shows the Doppler velocity and
the corresponding longitudinal component of the vector magnetic field in
the active regions NOAA~11158 on 2011 February 14 and NOAA~12266 on 2015
January 19 by the Helioseismic and Magnetic Imager (HMI) on board the
{\em Solar Dynamics Observatory} ({\em SDO}).
To obtain a representative nearly stationary pattern, we averaged over
a continuous series of Dopplergrams observed during 20~minutes.
The contribution from the five-minute oscillation is thus basically removed.
However, projection effects have not been compensated for.
A prominent Evershed flow can be seen in the strong magnetic structures
of NOAA~11158 (S20W17) due to its location south-west of disk center.
We can see a pattern of small-scale velocity nearby the active region.
A similar situation can also be found in the active region NOAA~12266
(S06E06) located near disk center.
These small-scale velocity patterns are indicated by
arrows in Figure~\ref{fig:corrmagheliveloa}.
Thus, the flow fields in Figure~\ref{fig:corrmagheliveloa} are expected
have contributions both from the active regions and the quiet Sun
in its proximity.

\begin{figure*}
\begin{center}
\includegraphics[width=.49\textwidth]{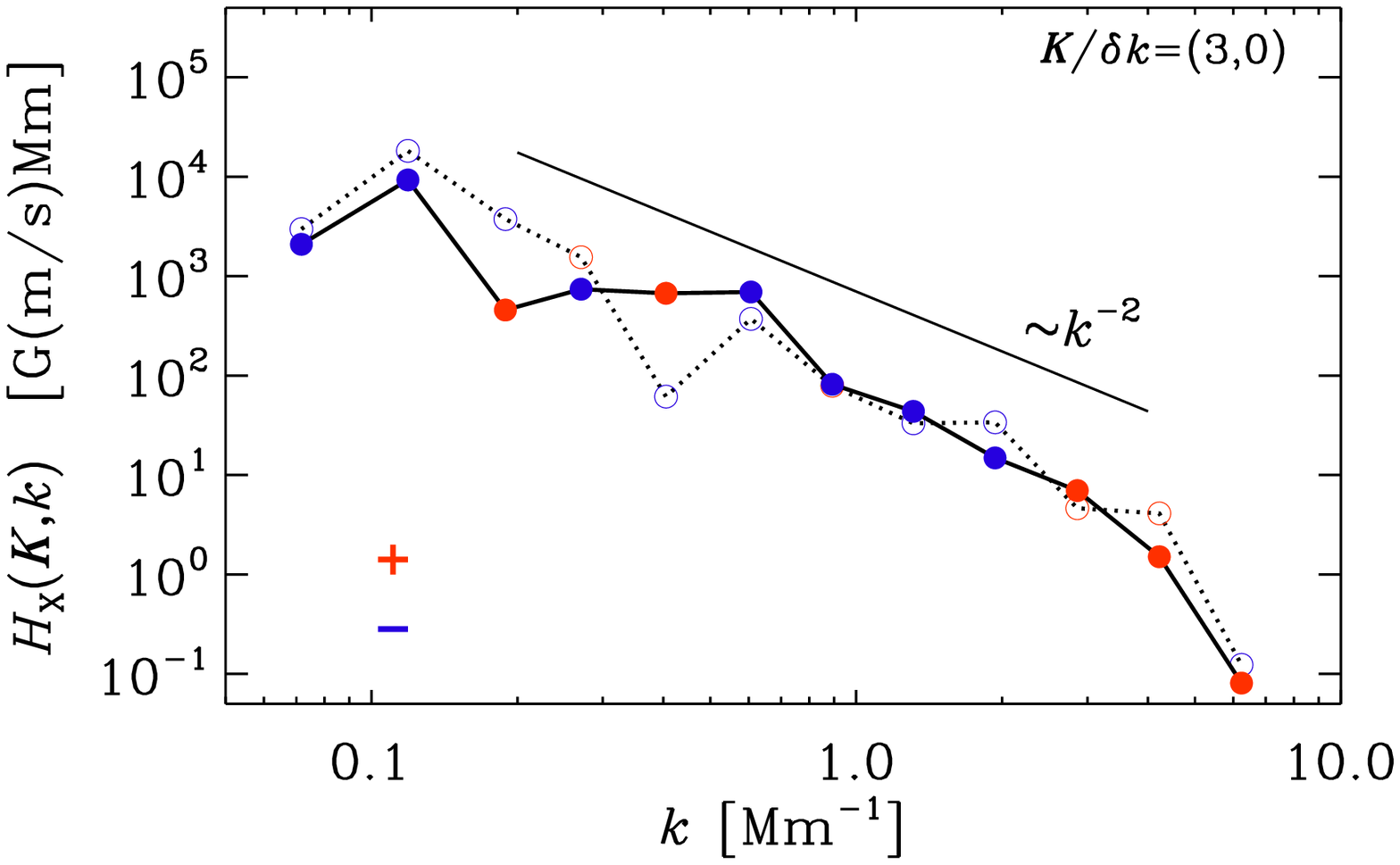}
\includegraphics[width=.49\textwidth]{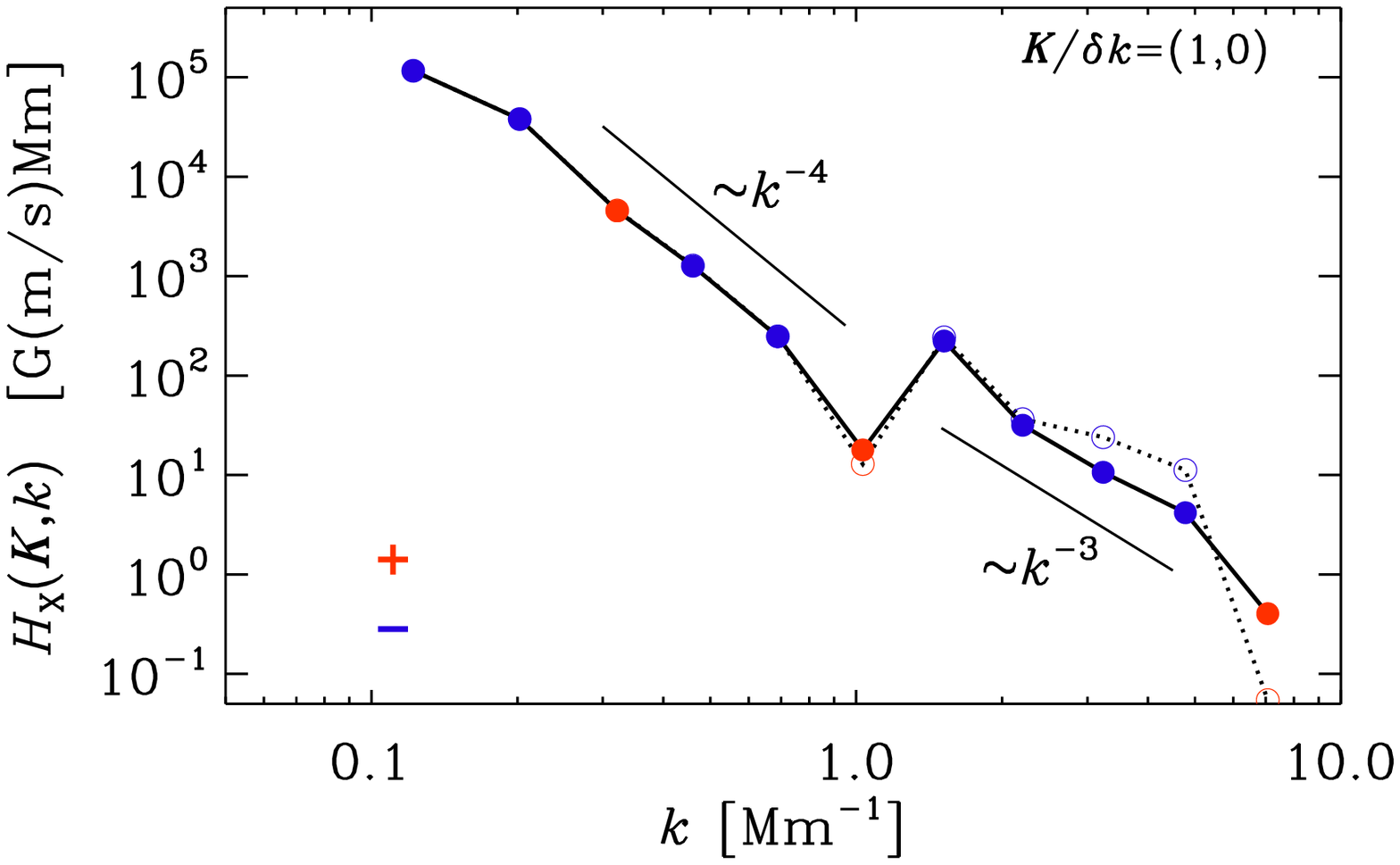}
\end{center}
\caption{{Cross helicity $H_{\rm X}(\KK,k)$ with two-scale}
analysis for active regions NOAA~11158 (left) and
NOAA~12266 (right) using $\KK/\delta k=(3,0)$ and $(1,0)$,
respectively; see \Eq{EKEMHXdx}.
{Red and blue symbols denote positive and negative values, respectively.}
Dotted lines with open symbols refer to strong fields only
(the line-of-sight magnetic field exceeds $\pm50\G$), while
solid lines with closed symbols apply to all points.
$\KK$ is given in units of $\delta k=2\pi/L$ defined below \Eq{EHvec}.
\label{fig:corrmaghelivelob_dx}
}\end{figure*}

Figure~\ref{fig:corrmaghelivelob} shows magnetic energy and scaled
helicity spectra, $E_{\rm M}(k)$ and $kH_{\rm M}(k)$, respectively,
inferred from the photospheric vector magnetograms of
NOAA 11158 \citep[cf.][]{Zhang2014} and 12266, along with the
corresponding kinetic energy spectra inferred from the Dopplergrams of
Figure~\ref{fig:corrmagheliveloa}.
{To reduce fluctuations in the helicity spectra, we have averaged
their values within logarithmically spaced wavenumber intervals.}
In the range of wavenumbers $k=0.5$--$2.5\Mm^{-1}$, the mean slopes of
$E_{\rm M}(k)$ and $k|H_{\rm M}(k)|$ are $ -1.8$ and $-3.4$, respectively,
for active region NOAA~11158, and $-1.5$ and $-2.1$, respectively,
for active region NOAA~12266.
The temporal variation of the slopes of $E_{\rm M}(k)$ and
$k|H_{\rm M}(k)|$ of active regions was discussed by \cite{Zhang2016}.
The green dashed lines $E_{\rm Kaq}(k)$ show kinetic energy spectra
of the two active regions and also those of the quiet sun.
A similar result was shown by \cite{ZhaoChou13} with the continuous
high spatial resolution Doppler observations of the Sun by {\em SDO}/HMI.
They determined the power from the convective flows in the $k\omega$
diagram and found that the location of the convective peak is shifted
toward lower wavenumbers in the power spectrum obtained from the sunspot
compared to that of the quiet Sun.

The green { solid}
lines $E_{\rm Ka}(k)$ in Figure~\ref{fig:corrmaghelivelob}
show { kinetic energy} spectra of the active region relative
to the magnetic structures with $|B_\||> 50\G$ only.
(The subscript ``a'' refers to active region.)
We find that the uprise of kinetic energy near $k=2$--$5\Mm^{-1}$ is now
removed for both active regions, and the slopes of the spectra of kinetic
energy are consistent with a $k^{-5/3}$ spectrum.
This removal is done by setting the velocity to zero at those points
where $|B_\||<50\G$ just prior to taking the Fourier transform.
In the range of wavenumbers $k=0.5$--$2.5\Mm^{-1}$, the mean slopes of
$E_{\rm Ka}$ are $-1.4$ for NOAA~11158 and $-1.5$ for NOAA~12266.
They are similar to
those of magnetic energy $E_{\rm M}(k)$ and scaled magnetic helicity
$kH_{\rm M}(k)$ in the photosphere in the range $k=1$--$2\Mm^{-1}$.
While for NOAA~11158 the slope of $E_{\rm Kaq}$ is $0.05$,
for NOAA~12266 it is $-0.2$ in the range $k=0.5$--$2.5\Mm^{-1}$.
(The subscript ``aq'' refers to the combination of active region and quiet Sun.)
The kinetic energy spectra $E_{\rm Kaq}(k)$ in the range $k=2$--$5\Mm^{-1}$
reflect the typical scale of the quiet Sun, which has
contributions from the granulation.

The $H_{\rm X}(k)$ spectra of \Eq{EKEMHX} have similar slopes as
$q\EK+\EM/q$ with $|H_{\rm X}|$ being about 10 times smaller than
the limit given by the total energy; see \Eq{Elimit}.
However, it has no definite sign.
This is different when taking the cancelation from the bipolarity
into account.
The two-scale method 
correlates functions whose wavenumbers differ by a small amount.
Consider as an example $v_\|=\sin k_+x$ and $B_\|=\cos k_-x$ with
$k_\pm=k\pm\delta k$ and $\delta k=\pi/d=2\pi/L$ being the lowest
wavenumber of the domain, then
\begin{equation}
2v_\|B_\|=\sin Kx + \sin2kx,
\label{modulation}
\end{equation}
with $K$ being the $x$ component of $\KK$,
has a low-wavenumber modulation proportional to $\sin Kx$ with
sign changes between the bipoles separated by $d$.
Comparing with \Figp{fig:corrmagheliveloa}{f} for NOAA~12266, where
$d\approx50\Mm$, the sign of $v_\|B_\|$ changes from positive values
for $x<0$ to negative ones for $x>0$.
This is the other way around than what is implied by the example
given in \Eq{modulation}.
Therefore, we expect $-{\rm Im}\,H_{\rm X}(\KK,k)$ itself to be negative.
This is indeed the case; see \Fig{fig:corrmaghelivelob_dx}, which
shows that $-{\rm Im}\,H_{\rm X}(\KK,k)$ is mostly negative for
NOAA~12266.

For NOAA~11158, there are two pairs of bipoles interlaced.
Each pair has an approximate separation $d\approx L/3$, but the
interlacing is not ideal and partially overlapping.
We tried $K_x/\delta k=1-3$, but none had as clean
a spectrum as NOAA~12266.
In \Fig{fig:corrmagheliveloa} we show for NOAA~11158 the spectrum for
$K_x/\delta k=3$, which had the least sign changes and is mostly
negative --- in broad agreement with a $\sin3\delta kx$ modulation in
\Figp{fig:corrmagheliveloa}{e}.
For NOAA~12266, on the other hand, $K_x/\delta k=1$ was found to
give the least sign changes -- in agreement with a $\sin\delta kx$
modulation in \Figp{fig:corrmagheliveloa}{f}.

A negative correlation between a large-scale field
proportional to $B_0\sin Kx$ and a correlation of the form
$\bra{v_\|B_\|}\approx-(\eta_{\rm T}/H_\rho)\sin Kx$
is theoretically expected \citep{RKB11}, where $\eta_{\rm T}$ is the
turbulent magnetic diffusivity and $H_\rho$ is the density scale height.
Using $\bra{v_\|B_\|}=\int H_{\rm X}(\KK,k)\,{\rm d}k\approx
-11,000\G\m\s^{-1}$ and $B_0\approx\Brms\approx300\G$, we have
$\bra{v_\|B_\|}/B_0\approx40\m\s^{-1}$ and $H_\rho=1\Mm$ yields
$\eta_{\rm T}\approx4\times10^{11}\cm^2\s^{-1}$, which is about
10 times less than what was found by \cite{RKS12}.

The slope for NOAA~12266 is between $-3$ and $-4$, which is much steeper
than that for NOAA~11158, where the slope was $-2$.
A steeper slope, especially at small $k$, is of interest when interpreting the $H_{\rm X}(\KK,k)$ spectrum
as an indicator for inverse cascading being a possible mechanism for
forming magnetic flux concentrations \citep{BGJKR14}.
However, there are sign changes at $k\approx0.3$ and $1\Mm^{-1}$,
which may not be compatible with this interpretation.
On the other hand, a jump similar to that at $k\approx1\Mm^{-1}$ has
also been seen in the simulations; see Figure~19 of \cite{BGJKR14}. 

\begin{figure}
\begin{center}
\includegraphics[width=\columnwidth]{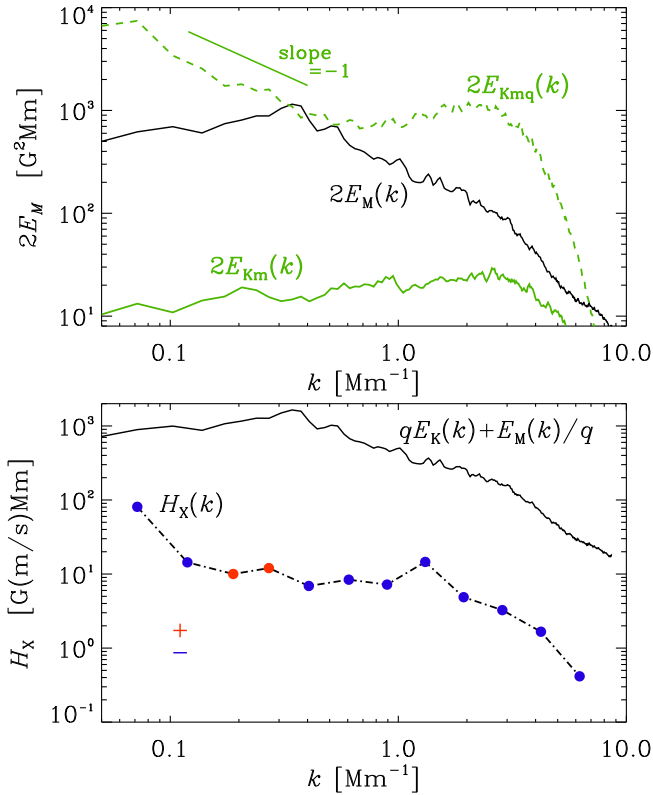}
\end{center}
\caption{
{The upper panel shows spectra}
of magnetic energy $E_{\rm M}(k)$ (black solid lines),
kinetic energy $E_{\rm Kmq}(k)$ (green dashed lines, for the whole
velocity in the field of view of the quiet Sun) and $E_{\rm Km}(k)$
(green solid
lines for the velocity related with magnetic features only).
{ The lower panel shows $qE_{\rm K}(k)+E_{\rm M}(k)/q$ (solid lines)
and $H_{\rm X}(k)$ (dashed-dotted lines; red and blue symbols denote positive
and negative values, respectively).} 
\label{fig:corrmagheliveloc}
\vspace*{6mm}
}\end{figure}

For comparison with the velocity field of active regions, we show in
Figure~\ref{fig:corrmagheliveloc} the kinetic energy spectra for the
velocity field of the quiet Sun near the center of the solar disk.
$E_{\rm Kmq}$ shows the spectrum of the whole velocity field in the field
of view of the quiet Sun, and it is almost consistent with the results
of \cite{ZhaoChou13}, while $E_{\rm Km}$ shows that of the velocity
field for magnetic fields with $|B_{\|}| >50\G$ only.
(Here the subscript ``mq'' refers to the whole region in the field of view
in the quiet sun.)
Due to the averaging over a series of continuous Dopplergrams observed
during 20~minutes, the contribution from the five-minute oscillation
has effectively been removed in our analyzed velocity field.
$E_{\rm M}(k)$ shows the spectrum of the magnetic energy inferred from
the longitudinal component of the magnetic field.
In the range of wavenumbers $k=0.5$--$2.5\Mm^{-1}$, the mean slope of
magnetic energy $E_{\rm M}$ is $-1.0$ and that of $E_{\rm Km}$ is $0.2$.
(Here the subscript ``m'' refers to the kinetic energy relative to areas
with magnetic field only.)
These are shallower than the $k^{-5/3}$ Kolmogorov spectrum.

\section{Conclusions}

{Our combined analysis of velocity and magnetic fields has shown
that, within active regions, kinetic and magnetic energy spectra have
similar slopes at intermediate scales.
Here the field is also close to maximally helical.
The magnetic helicity spectra of \cite{Zhang2014,Zhang2016} are
found to be identical to those composed of just $A_\|$ and $B_\|$.
This is analogous to the similarly constructed current helicity,
$\bra{J_\| B_\|}$, which is frequently employed in solar physics.
The helicity spectra are gauge-independent owing to the assumed
horizontal periodicity and independence of $z$.
This assumption affects only the smallest wavenumbers.
Unlike $\bra{J_\|B_\|}$, which captures helicity effects only on small
scales or high wavenumbers, here we have access to the
helicity decomposition into different wavenumbers.
}

{
Quite analogously, we have constructed cross helicity spectra.
Their signs switch with the sign of the mean vertical magnetic field,
which is the reason we have adopted here the two-scale approach.
This approach is familiar from mean-field dynamo theory \citep{RS75}
and has recently been applied to solar magnetic helicity spectra
\citep{BPS17,Singh18}, but it is the first time that it has been
applied to cross helicity.
It allows us to capture properties of global spectra, avoiding
cancellation from the different polarities of bipolar regions.
This approach worked particularly well for NOAA~12266, where the
separation of the two polarities is about half the extent of the
magnetogram.
By contrast, in NOAA~11158, two pairs of polarities are interlaced, making
the direct application of the two-scale approach less straightforward.
For NOAA12266, the spectral slope of the cross helicity is found to
be $-4$.}
A steep slope is suggestive of an inverse cascade phenomenon of cross
helicity \citep{BGJKR14} and can be a possible mechanism responsible
for causing magnetic flux concentrations into spots.
Further work incorporating a larger sample of active regions and a global analysis
would be an important future extension of this work.

\begin{acknowledgements}
This study is supported by grants from the National Natural Science
Foundation (NNSF) of China under the project grants 11673033, 11427803,
11427901 and Huairou Solar Observing Station of National Astronomical
Observatories, Chinese Academy of Sciences.
This work has been supported in part by
the NSF Astronomy and Astrophysics Grants Program (grant 1615100),
and the University of Colorado through its support of the
George Ellery Hale visiting faculty appointment.
\end{acknowledgements}

\end{document}